\documentclass[12pt]{article}

\usepackage{graphicx}

\newcommand{\beq}{\begin{equation}}
\newcommand{\eeq}{\end{equation}}
\newcommand{\beqa}{\begin{eqnarray}}
\newcommand{\eeqa}{\end{eqnarray}}
\newcommand{\beqar}{\begin{eqnarray*}}
\newcommand{\eeqar}{\end{eqnarray*}}

\begin{document}
\thispagestyle{empty}

\vspace*{-2mm}

\begin{center}
{\textbf{\Large Multi-photon annihilation of monopolium}}

\vspace{1cm}

H. Fanchiotti$^1$, C.A.~Garc\'\i a Canal$^1$ and V.~Vento$^2$
\vspace{12pt}

\textit{$^1$IFLP/CONICET and Departamento de F{\'\i}sica}\\ 
\textit{Universidad Nacional de La Plata, C.C.67, 1900, 
La Plata, Argentina}\\
\vspace{10pt}
\textit{$^2$Departamento de F{\'\i}sica Te\'orica-IFIC. Universidad de Valencia-CSIC.}\\ 
\textit{E-46100, Burjassot (Valencia), Spain}
\end{center}

\vspace{1cm}



\vspace{1cm}

\begin{abstract}
We  show that due to the large coupling constant of the monopole-photon interaction the annihilation  of monopole-antimonopole  and monopolium into many photons must be considered experimentally. For monopole-antimonopole annihilation and lightly bound monopolium, even in the less favorable scenario, multi-photon events (four and more photons in the final state) are dominant, while for strongly bound monopolium, although two photon events are important, four and six photon events are also sizable.

\end{abstract}

Inspired by an old idea of Dirac and Zeldovich \cite{Dirac:1931kp,Dirac:1948um,Zeldovich:1978wj} we proposed that monopolium, a bound state of monopole anti-monopole \cite{Zeldovich:1978wj,Hill:1982iq} could be easier to detect experimentally than free monopoles \cite{Vento:2007vy,Epele:2007ic}. We have already studied the annihilation of monopolium into two photons \cite{Epele:2012jn,Epele:2016wps}. In here we are going to  show that monopolium might annihilate most preferably into many low energy photons \cite{Epele:2016wps,Barrie:2016wxf}. This result motivates experimental searches of monopolium and  monopole-antimonopole by looking into multi-photon decays.

In our description of monopolium decays we follow closely that of positronium decays with two differences, namely the huge coupling constant in the magnetic case and  the dependence on the binding energy. The subtleties associated with the binding potential  \cite{Epele:2007ic,Barrie:2016wxf}  are of no relevance for the present estimation. 

The two and four photon decay channels of the ground state parapositronium have been studied in QED. In particular, the two photon channel is known up to $O(\alpha^3 \log ^2\alpha)$ \cite{Penin:2003jz,Lepage:1983yy,Khriplovich:1990eh,Czarnecki:1999mt} and  the four photon decay has been studied up to order $O(\alpha$) \cite{Adkins:1998ap},  where $\alpha$ is the fine structure  constant.  We show for the ratio of these channels the result to Leading Order  \cite{Billoire:1978wq,Muta:1982hb}, 

\begin{equation}
\frac{\Gamma_4}{\Gamma_2} = 0.277 \left(\frac{\alpha}{\pi}\right)^2  .
\label{ratio42QEDtrue}
\end{equation}

The factor in front of the coupling constant, $ 0.277$, contains the contribution of the $4!$ diagrams of the four photon amplitude and the $2!$ diagrams of the two photon one, to lowest order. The binding energy is  very small, a few eV, and has been neglected in the calculation,  therefore the energy factors cancel in the ratio. 

Let us for the sake of argument increase the photon coupling in Eq.(\ref{ratio42QEDtrue}) which leads to an increase in the four particle ratio. This increase in the ratio motivates the present investigation.

Let us assume that the monopole-photon coupling is analogous to the electron-photon coupling except for an effective vertex characterized by the dressed monopole magnetic charge  $g$ \cite{Zwanziger:1970hk}. Thus we extend the positronium calculation to monopolium just by changing $e\rightarrow g$. Recalling the parapositronium calculation in terms of the coupling we get,

\begin{equation}
\frac{\Gamma_4}{\Gamma_2} \sim F_{42} \left(\frac{\alpha_g}{\pi}\right)^2 \; \;\cdots \; \; \frac{\Gamma_{2n}}{\Gamma_2} \sim F_{2n2}\left(\frac{\alpha_g}{\pi}\right)^{2n-2},
\end{equation}

\noindent where $\alpha_g = \frac{1}{4 \alpha} \sim \frac{137}{4} \sim 34.25$ obtained from Dirac's Quantization Condition (DQC) \cite{Dirac:1931kp,Dirac:1948um}. The $F$'s represent the  contribution of all the Feynman amplitudes to the process shown as subindex after extracting the contribution of the magnetic charge, which is explicitly shown. For example, to leading order, $F_{42} \approx 0.277$ as seen in Eq(\ref{ratio42QEDtrue}).  We  perform the calculations, as is customary done in monopole physics, to leading order. Due to the large magnetic coupling the calculations can give at most a qualitative indication of magnitudes, which is what we can pursue at present. At the end of our analysis we comment on how non perturbative effects might affect our calculation.

\begin{figure}[ht]
\centerline{\includegraphics[scale= 1.0]{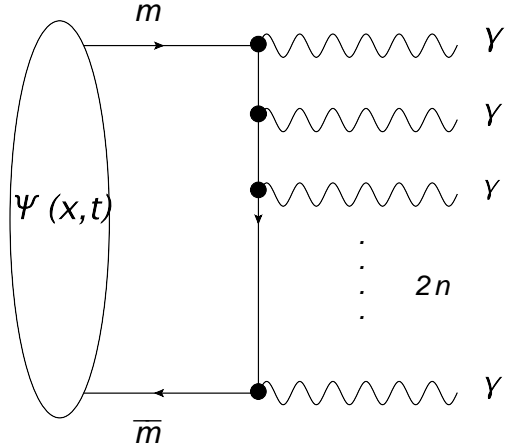}}
\vskip 0.3cm
\caption{One of the $2n!$ multi-photon emission diagrams.}
\label{multiphoton}
\end{figure}

In Fig. \ref{multiphoton} we show one of the $2 n!$ contributions to the amplitude for a $2n$ photon  decay to leading order, and we note that this type of contributions in the above ratios are determined only by vertices and propagators. The calculation for high $n$ with $2n!$ diagrams is out of the scope of any study. 
Let us discuss  first an educated estimation for large $n$. In the rest frame of the bound system the annihilation into many photons leads to an average momentum for each photon  much smaller than the mass of monopolium  and therefore much smaller than the mass of the monopole. In order to make an estimation of the above ratios we consider that in the propagators the monopole mass dominates over the momentum and therefore the calculation of the width, in units of monopole mass, depends exclusively on three factors: the number of diagrams $(2 n)!$, the photons' symmetry factor $1/(2 n)!$ and the phase space of the outgoing massless particles, namely

\begin{figure}[htb]
\centerline{\includegraphics[scale= 0.8]{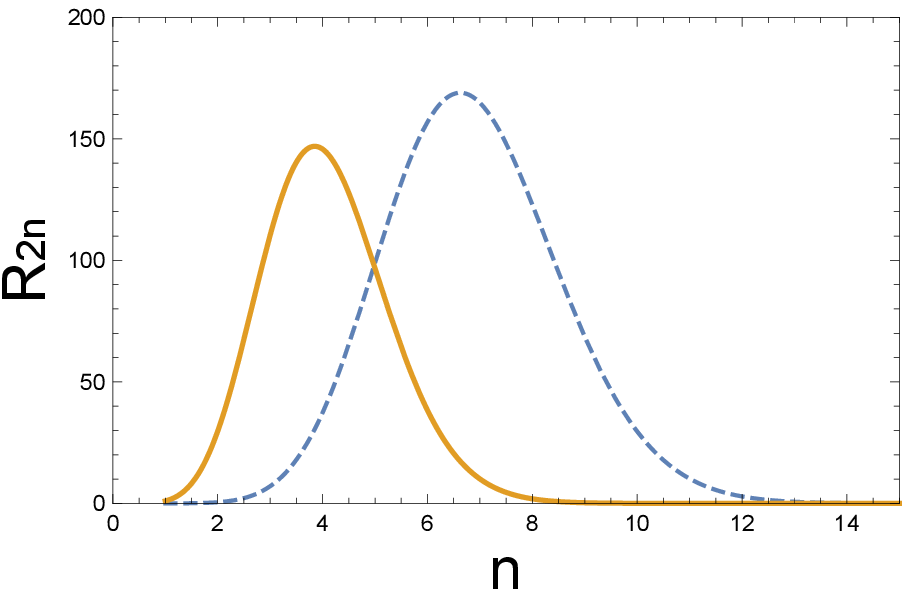} \hskip 1.0cm  \includegraphics[scale= 0.8]{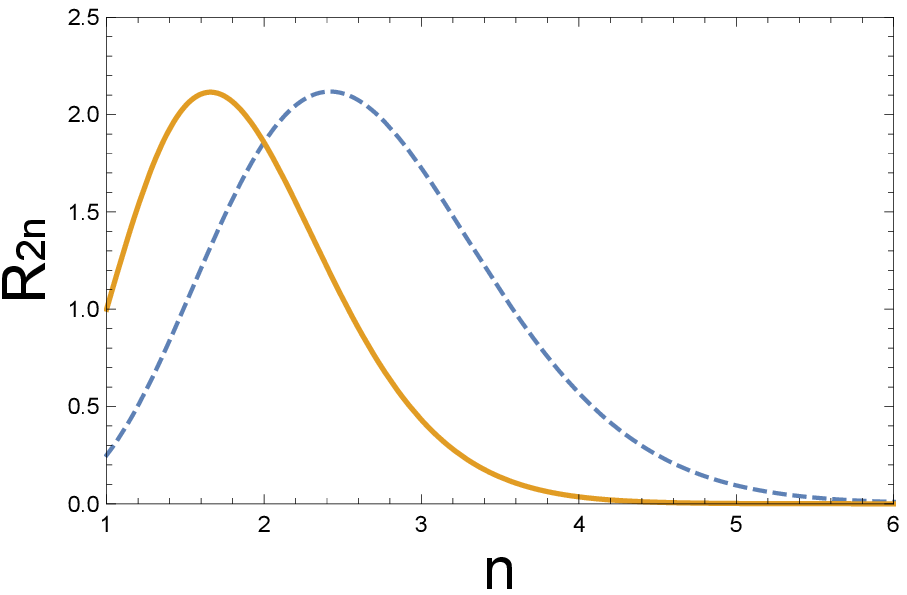}}
\vskip 0.3cm
\caption{The $\Gamma _{2n}/ \Gamma _2$  ratio as a function of $n$ calculated according to Eqs. (\ref{ratio}) and (\ref{ratiofactor}). The left figure is for zero binding energy and the right figure for $E_b \sim m$. The solid curves represent the calculation with the interference factor and the dashed curves the one without the factor. In order to have the two curves at the same scale the no interference ratios had to be divided by 250 left and by 4 right.}
\label{ratioN2}
\end{figure}

\begin{equation}
(phsp)^{2n}= \frac{1}{2} \frac{1}{(4 \pi)^{4n-3} }\frac{M^{4n-4}}{\Gamma (2n) \Gamma (2n-1)},
\end{equation}

\noindent where $M$ is the monopolium mass and $n = 1,2 ,3, \ldots$ being $2n$ the number of photons emitted.

With all these approximations we obtain the expression

\begin{equation}
\frac{\Gamma_{2n}}{\Gamma_2}  = \left(\frac{\alpha_g}{\pi}\right)^{2n-2}  \left(\frac{M}{2m}\right)^{4n-4} \frac{2n!}{2! (2n-1)! (2n-2)!}.
\label{ratio}
\end{equation}
Note that this equation leads to $\Gamma_{2}/\Gamma_2 =1$ and for $n=2$  and $M=2m$, one recovers the parapositronium case, $\Gamma_{4}/\Gamma_2 = \left(\frac{\alpha_g}{\pi}\right)^{2}$, with the interference factor missing (recall Eq.(\ref{ratio42QEDtrue})).  In order to incorporate this effect we make a second estimate. In the first estimate we have assumed $p^2$ to be very small compared with $m^2$ in the propagator an approximation valid for large $n$. Let us assume for the second estimate that on the contrary $p^2 \sim m^2$, an approximation which might be adequate for small $n$. This aproximation introduces  in Eq.(\ref{ratio}) a factor $(\frac{1}{2})^{2n-2}$ leading to

\begin{equation}
\frac{\Gamma_{2n}}{\Gamma_2}  = \left(\frac{1}{2}\right)^{2n-2} \left(\frac{\alpha_g}{\pi}\right)^{2n-2}  \left(\frac{M}{2m}\right)^{4n-4} \frac{2n!}{2! (2n-1)! (2n-2)!}.
\label{ratiofactor}
\end{equation}
For $n=2$ this factor is $0.25$ which is very close to true calculation to leading order $0.277$.  This factor is extremely suppressing for large $n$, where the approximations discussed initially might be better. We show results with and without this factor to determine a region of confidence. If the leading order calculation were all there is, the true result would be between these two limiting expressions. We discuss possible non perturbative effects at the end of the analysis. In our expressions we consider the effect of the binding energy not taken into account in the conventional positronium analysis.

\begin{figure}[htb]
\centerline{\includegraphics[scale= 1.0]{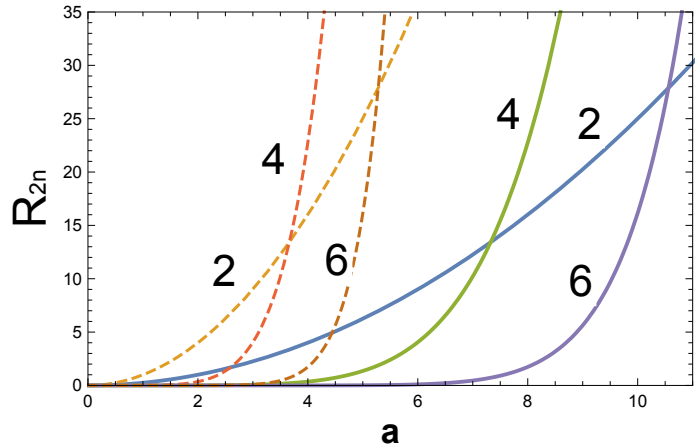}}
\vskip 0.3cm
\caption{The $\Gamma _{2n}/ \Gamma _2$  ratio as a function of $a$ calculated according to Eqs. (\ref{ratio}) and (\ref{ratiofactor}) for n = 2, 4, 6 and zero binding energy. The solid curves represent the calculation with the interference factor and the dashed curves the one without the factor.}
\label{GnG2E0}
\end{figure}

In Fig.\ref{ratioN2} we plot Eqs.(\ref{ratio}) and (\ref{ratiofactor}) for two different binding energies and we get  bell shape distributions. For small binding energies ($M\approx 2m$) the value of $n$ on the average is  $ \overline{n} \sim 7$ with a deviation of $\Delta n \sim \pm 2$. For large binding energies ($M\sim m$) the average value of $n$ is $ \overline{n} \sim 3$ with a deviation of $\Delta n \sim \pm 1$. If the interference factor is included the multiplicity ($2\overline{n}$) is reduced to $ \overline{n} \sim 4$ with $\Delta n \sim \pm 1$ for small binding and $ \overline{n} \sim 2$ with $\Delta n \sim \pm 1$ for large binding. Thus the multiplicity decreases as the binding energy increases. However, even with the strongly suppressing interference factor included, four photon emission is favored.

Since the above curves do not provide a quantitative  estimate of the increase in the ratio we show in Fig. \ref{GnG2E0} some ratios as a function of $a$ for a final state of 4, 8 and 12 photons  with interference factor (solid) and without factor (dashed) for very small binding energy. This case of small binding energy corresponds very closely to monopole-antimonopole annihilation. The effect is large even with the interference factor included. Note that the monopole coupling corresponds to $a \sim 11$.

\begin{figure}[ht]
\centerline{\includegraphics[scale= 1.0]{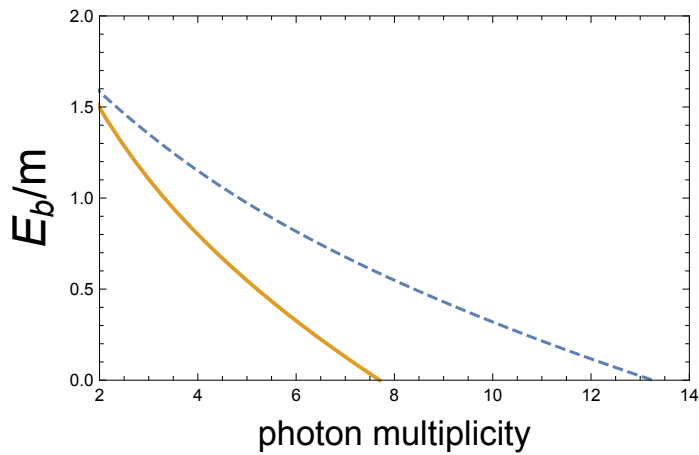}}
\vskip 0.3cm
\caption{Binding energy in units of monopole mass as a function of average photon multiplicity in the annihilation. The solid curve represents the calculation with interference factor while the dashed curve is that without that factor.}
\label{ratiobindingexact}
\end{figure}

We now study the dependence of multiplicity with the binding energy. To do so we find the maximum of the ratio in Eqs. (\ref{ratio}) and (\ref{ratiofactor}) as a function of binding energy. The result is plotted in Fig.(\ref{ratiobindingexact}) were we show the  binding energy in units of monopole mass as a function of the average photon multiplicity in the annihilation. The outcome is clear,  large average multiplicities, 8-12 photons, arise if the binding energy is small, $M \sim 2m$, while smaller average multiplicities, 4-6 photons,  occur for large bindings, $M\sim m$. In the latter case considerable rates extend  up to multiplicities of 8 photons as is shown in Fig. \ref{ratioN2} right.

\begin{figure}[htb]
\centerline{\includegraphics[scale= 1.0]{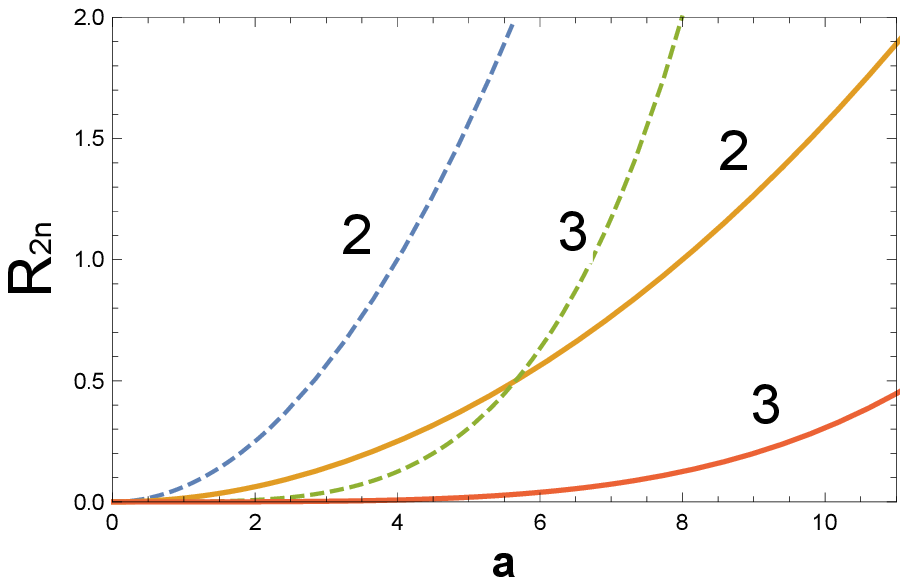}}
\vskip 0.3cm
\caption{The $\Gamma _{2n}/ \Gamma _2$  ratio as a function of $a$ calculated according to Eqs. (\ref{ratio}) and (\ref{ratiofactor}) for n = 2, 3 and large binding energy ($M \sim m$). The solid curves represent the calculation with the interference factor and the dashed curves the ones without that factor.}
\label{GnG2E1}
\end{figure}

In order to see  quantitatively the increase in the ratios  for large binding energy ($M\sim m$) we plot in Fig \ref{GnG2E1} as a function of $a$ the ratios for 4 and 6 photons calculated according to Eqs. (\ref{ratio}) and (\ref{ratiofactor}) . In this case the rates are smaller and also the multiplicities as  seen in Fig. \ref{ratiobindingexact}. It is important to realize that the binding energy effect is very suppressing in the phase space formula. However, if the mass of the monopole is large ($> 1$ TeV), binding energies as large as its mass are not to be expected.

We can get an analytic formula for the most probable photon decay channel by calculating the maximum of the  logarithm of Eqs. (\ref{ratio}) and (\ref{ratiofactor}) using Sterling's formula 
($k! \sim k^k e^{-k} \sqrt{2 \pi k}$). Sterling's formula is quite good even for low $k$, i.e for $k=2$  it gives $1.919$, for $k=3$, $5.836$, and $k=4$, $23.506$. Thus we can consider the equations we are going to derive good approximations for any $n$.
Let us write a generic interference factor $(\delta)^{n-1}$ in front of Eq. (\ref{ratio}) which for $\delta=1$ gives Eq.(\ref{ratio}) itself and for $\delta=0.25$ Eq.(\ref{ratiofactor}). The values of $n$ for the maximum decay rates are given by 
the solutions of  the equation

\begin{equation}
 n = \frac{ \alpha_g  \sqrt{\delta}}{2 \pi} \left(1- \frac{E_b}{2m}\right)^2 \;  \exp({\frac{1}{4n}}) + 1,
 \label{max}
\end{equation}
for  specific values of $\delta$ and the binding energy $E_b$. For large $n$, which occurs for small binding energy we get the approximate solution

\begin{equation}
n  \approx  \frac{ \alpha_g \sqrt{\delta}}{2 \pi} +1,
\end{equation}

\noindent which is very illuminating because it shows explicitly the effect of the coupling constant in increasing the photon multiplicities as seen numerically in Fig. \ref{ratioN2}.

All the  approximations used thus far are  valid for Dirac's original formulation. Some coupling schemes lead to small effective couplings close to threshold \cite{Epele:2012jn,Epele:2016wps,Milton:2006cp}.  These velocity dependent schemes proceed by changing $g \rightarrow \beta g$, where

\begin{equation}
\beta = \sqrt{1-\frac{M^2}{s}},
\end{equation}
$M$ being the mass of monopolium and $s$ the center of mass energy of the process. In the case of monopole-antimonopole production $M \rightarrow 2 m$, where $m$ is the mass of the monopole (antimonopole).  Thus, in these schemes, all photon widths vanish at threshold.
Close to threshold two photon decays are dominant, since the ratio (Eqs. (\ref{ratio}) and (\ref{ratiofactor}))  acquire a  factor $\beta^{4n-4}$.
However, by looking at the approximate solution which is now modified to

\begin{equation}
n  \approx  \frac{ \alpha_g \beta^2 \sqrt{\delta}}{2 \pi} +1.
\end{equation}
one can understand what happens. Close to threshold the two photon decay is the dominant process but given the size of the coupling for $\beta > 0.5$  multi-photon decays start to be important. Given that in most processes studied $\beta$ rises rapidly  \cite{Epele:2012jn,Epele:2016wps} our present analysis holds slightly away from threshold.

\begin{figure}[htb]
\centerline{\includegraphics[scale= 1.0]{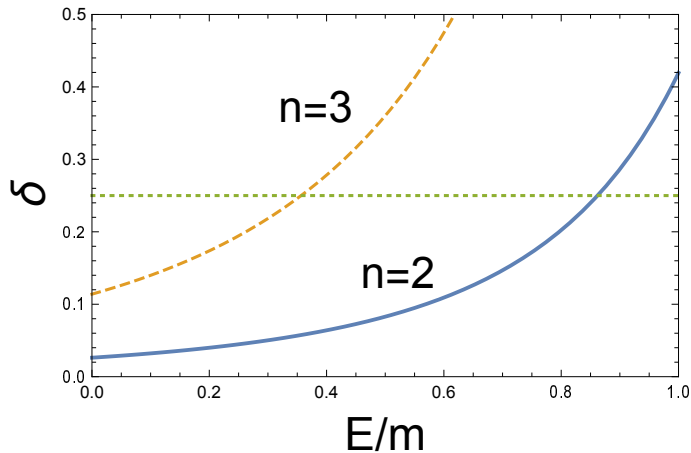}}
\vskip 0.3cm
\caption{The interference factor $\delta$ as a funtion of binding energy for four and six photon decays according to Eq. (\ref{max}).}
\label{nonperturbative}
\end{figure}

Finally, we would like to make some comments about non perturbative effects. Given the large value of the coupling constant it is evident that our calculation is merely qualitative and aimed at proposing new signals to discover monopoles at lower photon energies. Let us assume for the following discussion  the worst possible scenario namely that non perturbative effects make  the multi-photon channels weaker. We parametrize the non perturbative effects  by  an effective $\delta$ . How small can $\delta$ get  to make four and six photon decay ratios irrelevant? We use Eq.(\ref{max}) for $n=2,3$ and plot $\delta$ as a funtion of binding energy in Fig. \ref{nonperturbative}.

From the figure it is apparent that a small bound monopolium  produces preferably multi-photon decays  up to very small interference factors. As the binding energy increases the possibility of multi-photon decays decreases.  Note however that the four photon decay is greater or comparable to two photon decays up to interference factors many times smaller than the one used in this calculation and note that monopole-antimonopole annihilation behaves much like zero binding monopolium.

Our  analysis shows that monopole-antimonopole annihilation or lightly bound monopolium decays lead preferably to  multi-photon  events. In particular one should look  for four or more photons in the final state. If monopolium is strongly bound the situation changes and although two photon rates are important it is also certain that four and six photon rates may be sizable. Thus in any circumstance one should aim at looking at four and six photon events with the multiplicity characterizing the dynamics of the binding.

Present searches of low mass monopoles have been carried out by making use of their magnetic properties trapped in matter, by looking for tracks associated with their ionization properties and by studying two photon decays at colliders \cite{Patrizii:2015uea,MoEDAL:2016jlb,Acharya:2016ukt}. At colliders like the LHC the conservation of magnetic charge implies that either  monopolium or a pair monopole-antimonopole could be produced. Monopolium, being chargeless,  is difficult to detect except for its decay properties. If monopole-antimonopole are produced close to threshold they might annihilate given their large interaction before reaching the detector. In this experimental scenario we might not have  been able to detect monopolium and/or monopoles because we have been looking into trapped monopoles, ionization remnants or the two photon channel, all of which according to our present investigation are less probable than four or more photon events. Our study shows that a characteristic feature of monopole-antimonopole annihilation and monopolium decay is to find more than two photons coming from the annihilation vertex. At threshold these photons have {\it on average} smaller energy than the energy of a typical collider process and the multiplicity is directly related to the strength of their interaction.

To conclude we state, that in view of the fact that the exact dynamics of monopoles and their properties are not available, large multiplicity of photon events  might be the signal for the discovery of these elusive particles. Experiments should be ready to incorporate this feature into their analysis.

\section*{Acknowledgments}
Vicente Vento acknowledges the hospitality extended to him by the IFLP/CONICET and Departamento de F{\'\i}sica de la Universidad Nacional de La Plata were this work was finalized. The work  has been partially supported by ANPCyT and CONICET of Argentina, and by MINECO (Spain) Grant. No. FPA2013-47443-C2-1-P, GVA-PROMETEOII/2014/066 and  SEV-2014-0398.


\begin{thebibliography}{99}

\bibitem{Dirac:1931kp}
  P.~A.~M.~Dirac,
  Proc.\ Roy.\ Soc.\ Lond.\  A {\bf 133} (1931) 60.
  
 \bibitem{Dirac:1948um}
  P.~A.~M.~Dirac,
  Phys.\ Rev.\  {\bf 74} (1948) 817.


\bibitem{Zeldovich:1978wj}
  Y.~B.~Zeldovich and M.~Y.~Khlopov,
  Phys.\ Lett.\  B {\bf 79} (1978) 239.
  
\bibitem{Hill:1982iq}
  C.~T.~Hill,
  Nucl.\ Phys.\ B {\bf 224} (1983) 469.
  doi:10.1016/0550-3213(83)90386-3

\bibitem{Vento:2007vy}
  V.~Vento,
  Int.\ J.\ Mod.\ Phys.\ A {\bf 23} (2008) 4023
  doi:10.1142/S0217751X08041669
  [arXiv:0709.0470 [astro-ph]].

\bibitem{Epele:2007ic}
  L.~N.~Epele, H.~Fanchiotti, C.~A.~Garcia Canal and V.~Vento,
  Eur.\ Phys.\ J.\ C {\bf 56} (2008) 87
  doi:10.1140/epjc/s10052-008-0628-0
  [hep-ph/0701133].


\bibitem{Epele:2012jn}
  L.~N.~Epele, H.~Fanchiotti, C.~A.~G.~Canal, V.~A.~Mitsou and V.~Vento,
  Eur.\ Phys.\ J.\ Plus {\bf 127} (2012) 60
  doi:10.1140/epjp/i2012-12060-8
  [arXiv:1205.6120 [hep-ph]].

\bibitem{Epele:2016wps}
  L.~N.~Epele, H.~Fanchiotti, C.~A.~G.~Canal, V.~A.~Mitsou and V.~Vento,
  arXiv:1607.05592 [hep-ph].



\bibitem{Barrie:2016wxf}
  N.~D.~Barrie, A.~Sugamoto and K.~Yamashita,
  PTEP {\bf 2016} (2016) no.11,  113B02
  doi:10.1093/ptep/ptw155
  [arXiv:1607.03987 [hep-ph]].

\bibitem{Penin:2003jz}
  A.~A.~Penin,
  Int.\ J.\ Mod.\ Phys.\ A {\bf 19} (2004) 3897
  doi:10.1142/S0217751X04020154
  [hep-ph/0308204].

\bibitem{Adkins:1998ap}
  G.S. Adkins and E.D. Pfahl
  Phys.\ ReV.\  D {\bf 59} (1998) R915.
  
\bibitem{Lepage:1983yy}
  G.~P.~Lepage, P.~B.~Mackenzie, K.~H.~Streng and P.~M.~Zerwas,
  Phys.\ Rev.\ A {\bf 28} (1983) 3090.
  doi:10.1103/PhysRevA.28.3090
  
\bibitem{Khriplovich:1990eh}
  I.~B.~Khriplovich and A.~S.~Yelkhovsky,
  Phys.\ Lett.\ B {\bf 246} (1990) 520.
  doi:10.1016/0370-2693(90)90641-I
  
\bibitem{Czarnecki:1999mt}
  A.~Czarnecki and S.~G.~Karshenboim,
  hep-ph/9911410.
  

\bibitem{Billoire:1978wq}
  A.~Billoire, R.~Lacaze, A.~Morel and H.~Navelet,
  Phys.\ Lett.\  {\bf 78B} (1978) 140.
  doi:10.1016/0370-2693(78)90367-2

\bibitem{Muta:1982hb}
  T.~Muta and T.~Niuya,
  Prog.\ Theor.\ Phys.\  {\bf 68} (1982) 1735.
  doi:10.1143/PTP.68.1735


\bibitem{Zwanziger:1970hk}
  D.~Zwanziger,
  Phys.\ Rev.\ D {\bf 3} (1971) 880.
  doi:10.1103/PhysRevD.3.880

 
\bibitem{Milton:2006cp}
  K.~A.~Milton,
  Rept.\ Prog.\ Phys.\  {\bf 69} (2006) 1637
  doi:10.1088/0034-4885/69/6/R02
  [hep-ex/0602040].
 
  
\bibitem{Patrizii:2015uea}
  L.~Patrizii and M.~Spurio,
  Ann.\ Rev.\ Nucl.\ Part.\ Sci.\  {\bf 65} (2015) 279
  doi:10.1146/annurev-nucl-102014-022137
  [arXiv:1510.07125 [hep-ex]].
  
\bibitem{Acharya:2014nyr}
  B.~Acharya {\it et al.} [MoEDAL Collaboration],
  Int.\ J.\ Mod.\ Phys.\ A {\bf 29} (2014) 1430050
  doi:10.1142/S0217751X14300506
  [arXiv:1405.7662 [hep-ph]].
  
\bibitem{MoEDAL:2016jlb}
  B.~Acharya {\it et al.} [MoEDAL Collaboration],
  JHEP {\bf 1608} (2016) 067
  doi:10.1007/JHEP08(2016)067
  [arXiv:1604.06645 [hep-ex]].
  
\bibitem{Acharya:2016ukt}
  B.~Acharya {\it et al.} [MoEDAL Collaboration],
  Phys.\ Rev.\ Lett.\  {\bf 118} (2017) no.6,  061801
  doi:10.1103/PhysRevLett.118.061801
  [arXiv:1611.06817 [hep-ex]].



\end{thebibliography}
\end{document}